# "An analysis of the classical Doppler Effect"[1] revisited

**Bernhard Rothenstein**

Politehnica University of Timisoara, Physics Dept., Timisoara, Romania

E-mail: bernhard_rothenstein@yahoo.com

**Corina Nafornita**

Politehnica University of Timisoara, Communications Dept., Timisoara, Romania

**Abstract**
After having shown that the formula which describes the Doppler effect in the general case holds only in the case of the "very high" frequency assumption, we derive free of assumptions Doppler formulas for two scenarios presented in the revisited paper.

1. **Introduction**

In a recent paper published in this Journal, the Authors[1] derive Doppler formulas for the classical acoustic Doppler Effect adapted to different scenarios. They extend to two space dimensions an approach proposed by Donges[2]. The purpose of our Note is to show that the Doppler formulas they derive hold only in the case of the "very small" period assumption. They are small enough in order to ensure that the moving source emits two successive wave crests (positive maximums) from the same point in space and that the observer receives them from another but the same point in space. We also derive for some of the scenarios considered by the Authors, free of assumptions Doppler formulas.

## 2. How general is the Doppler formula in the general case?

We present the following scenario in Figure 1. It involves a source S and a receiver R, in arbitrary motion relative to an inertial reference frame K(XOY). $P_S$ and $P_R$ represent the trajectories of S and R respectively. $S(r_S, \theta_S)$ represents an instantaneous position of S from where it emits wave crests when a clock $C_S(\vec{r_S})$ located in front of him reads $t_S$. R receives the wave crest at an instantaneous position $R(\vec{r_R})$ when a clock $C_R(\vec{r_R})$ located there reads $t_R$. The clocks of the K(XOY) are synchronized and so they display at each time the same running time. If $\vec{U}$ stands for the wave velocity, it is obvious that

$$\vec{U}(t_R - t_S) = \vec{r_R} - \vec{r_S}. \tag{1}$$

Deriving both sides of Eq.(1) with respect to $t_R$ and taking into account that by definition

$$\vec{v}_{S,r} = \frac{d\vec{r_S}}{dt_S} \tag{2}$$

and

$$\vec{v}_{R,r} = \frac{d\vec{r_R}}{dt_R} \tag{3}$$

represent the instantaneous radial components of the velocities of S and R respectively, we obtain

$$\vec{U}\left(1 - \frac{dt_S}{dt_R}\right) = \vec{v}_{R,r} - \vec{v}_{S,r}\frac{dt_S}{dt_R} \tag{4}$$

Using the dot product for vectors, Eq.(4) leads to

$$U^2\left(1 - \frac{dt_S}{dt_R}\right) = \vec{U}\cdot\vec{v}_{R,r} - \vec{U}\cdot\vec{v}_{S,r}\frac{dt_S}{dt_R} \tag{5}$$

Solved for $\frac{dt_S}{dt_R}$, Eq.(5) leads to

$$\frac{dt_S}{dt_R} = \frac{1 - \frac{v_{R,r}}{U}\cos\theta_R}{1 - \frac{v_{S,r}}{U}\cos\theta_S} \tag{6}$$

where $\theta_R$ and $\theta_S$ represent the angles made by $\vec{v}_{R,r}$ and $\vec{U}$ and by $\vec{v}_{S,r}$ and $\vec{U}$ respectively. As we see, Eq.(6) relates to "very small" changes in the readings of clocks $C_S(\vec{r_S})$ and $C_R(\vec{r_R})$, small enough in order to consider that during the measurement of $dt_S$ and $dt_R$ they remain at the same point in space. If we consider that $dt_S$ represents the "very small" period at which the source emits successive wave crests and that $dt_R$ represents the "very small" period at which R receives them, we can consider that Eq.(6) represents a Doppler formula which holds only in the case of the "very small" period assumption. Expressed

as a function of the "very high" frequencies $f_S = (dt_S)^{-1}$ and $f_R = (dt_R)^{-1}$, Eq.(6) becomes

$$f_R = f_S \frac{1 - \frac{v_R}{U}\cos\theta_R}{1 - \frac{v_S}{U}\cos\theta_S} \tag{7}$$

Eq.(7) is the starting point in the revisited paper[1]. The Authors fail to mention the limits under which it holds. Similar Doppler formulas are derived[3] considering that two observers S and R move in the space through which a plane wave propagates. Such Doppler formulas hold only in the case when the distance between the source that generates the plane wave and the involved observers is "very big".

The way in which we derived Eq.(7) shows that the time interval during which we measure the instantaneous velocity and the period of the oscillations in the wave are of the same order in magnitude.

### 3. Free of assumptions approach to the Doppler Effect

#### 3.1. Source S is at rest and observer R moves with constant acceleration along the direction, which joins S and R.

The scenario involves a stationary source S and an observer R in uniform accelerating motion with acceleration $a$. At the origin of time, R is located in front of the source and its instantaneous velocity at a time $t$ is

$$v = at \tag{8}$$

and its instantaneous position relative to the source is defined by the space coordinate

$$x = \frac{at^2}{2} \tag{9}$$

if we consider that the trip of R starts at $t = 0$ from the location of the source.

In the case of the "very small" period assumption, when we consider that R could perform a continuous recording of the frequency $f_R$, we use Eq.(7) making $v_S = 0$, $\theta_R = 0$, $v_R = at$ and we recover the formula proposed in paper[1]

$$f_R = f_S \left(1 - \frac{at}{U}\right). \tag{10}$$

In a free of assumptions approach, we consider that S emits a first wave crest at $t = 0$ which is instantly received by R. The $n$-th wave crest is emitted at $nT$ ($T$ is the emission period) and R receives it at a time $t_n$. Equating the distance traveled by R between the reception of the first and of the $n$-th wave crest with the distance traveled by the $n$-th wave crest we obtain

$$\frac{at_n^2}{2} = U(t_n - nT) \tag{11}$$

resulting

$$t_n = \frac{U}{a}\left(1 - \sqrt{1 - \frac{2a}{U}nT}\right) \qquad (12)$$

The time interval between the receptions of two successive wave crests is

$$T_{n-1,n} = t_n - t_{n-1} = \frac{2T}{B_{n-1} + B_n} \qquad (13)$$

where

$$B_n = \sqrt{1 - \frac{2a}{U}nT} \ . \qquad (14)$$

Eq.(12) enables us to define a Doppler factor

$$D = \frac{f_R}{f_S} = \frac{B_{n-1} + B_n}{2} \qquad (15)$$

and we present in Figure 2 its variation with $n$ for $U = 340\,\text{m/s}$ and $a = 10\,\text{m/s}^2$ and different values of the frequency $f_R = 1/T$ in the acoustic domain. When the velocity of R equates the wave velocity, he loses the contact with the source.

### 3.2. The observer moves with constant velocity

The Authors of the revisited paper[1] consider the case when S is at rest at the origin O of K(XOY) and R moves with constant velocity v in perpendicular direction to the OX axis (Figure 3). In a free of assumptions approach, R$_0$(d,0) represents a position of R at $t = 0$ where he receives a wave crest emitted at $t = -\frac{d}{U}$. R$_n$(d,v$t_n$) represents an instantaneous position of R where he receives at $t_n$ the wave crest emitted at $-\frac{d}{U} + nT$. Pythagoras' theorem applied to Figure 3 leads to

$$\frac{d^2}{U^2} + \frac{v^2}{U^2}t_n^2 = \left(t_n + \frac{d}{U} - nT\right)^2 \qquad (16)$$

Solved for $t_n$ Eq.(16) leads to

$$t_n = T\frac{n - \frac{df}{U} + \sqrt{\frac{d^2}{U^2}f^2 + \frac{v^2}{U^2}n\left(n - \frac{2d}{U}f\right)}}{1 - \frac{v^2}{U^2}} \qquad (17)$$

The time interval between the receptions of two successive wave crests is

$$T_{n-1,n} = t_n - t_{n-1} = T\frac{1 + B_n - B_{n-1}}{1 - \frac{v^2}{U^2}} \qquad (18)$$

where

$$B_n = \sqrt{\frac{d^2}{U^2}f^2 + \frac{v^2}{U^2}n\left(n - \frac{2d}{U}f\right)} \qquad (19)$$

and $f_S = T^{-1}$.

Eq.(18) enables us to define a Doppler factor D given by

$$D = \frac{f_R}{f} = \frac{T}{T_{n-1,n}} = \frac{1 - \frac{v^2}{U^2}}{1 + B_n - B_{n-1}} = \frac{1 - \frac{v^2}{U^2}}{1 + \frac{v^2}{U^2} \frac{2n - 1 - \frac{2d}{U}f}{B_n + B_{n-1}}} \qquad (20)$$

We present in Figure 4 the variation of the Doppler factor $D$ with $n$ for $v/U = 0.8$ (subsonic velocity) and with $f$ as a parameter in the acoustic range of frequencies. For $n \to \pm\infty$, Eq.(20) leads to the formula which accounts for the longitudinal Doppler Effect

$$D_{n \to \pm\infty} = 1 \mp \frac{v}{U} \qquad (21)$$

## 4. Conclusions

The free of assumptions approach to the classical Doppler Effect reveals some of its properties, obscured in many cases. They are:
    In the time interval between the receptions of two successive wave crests, the observer R has not enough information in order to reckon the reception period,
    The Doppler factor is not linear as it depends not only on the relative velocity of observer relative to the source but also on the frequency of the source.

**References**

[1] C.Neipp et al., "An analysis of the classical Doppler effect," 2003 Eur.J.Phys. 24, 497-505
[2] A.Donges, "A simple derivation of the acoustic Doppler shift formulas," Eur.J.Phys. 23, L25-L28, 2002
[3] C.Moller, "The Theory of Relativity," Clarendon Press Oxford 1972 Ch.2.9

**Figure Captions**
**Figure 1**. Scenario that leads to the "general" Doppler formula.
**Figure 2**. The variation of the Doppler factor with the order number of the received wave crest, in the case of a longitudinal Doppler Effect experiment, involving a stationary source and uniformly accelerating receiver.
**Figure 3**. Doppler Effect experiment, with stationary source and observer, who moves with constant velocity, parallel to the OY axis.
**Figure 4**. Variation of the Doppler factor, with the order number of the received wave crest, in the case of the scenario presented in Figure 3.

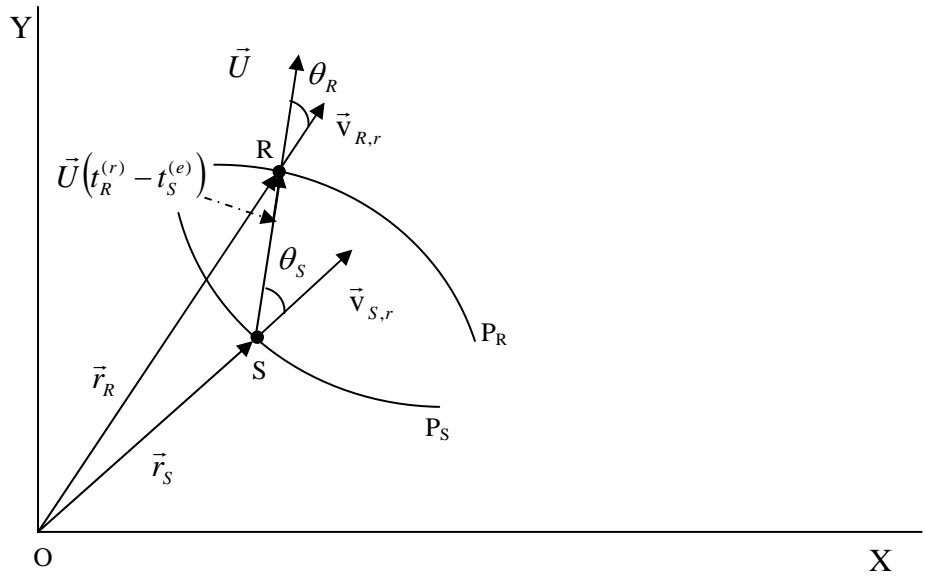

**Figure 1**

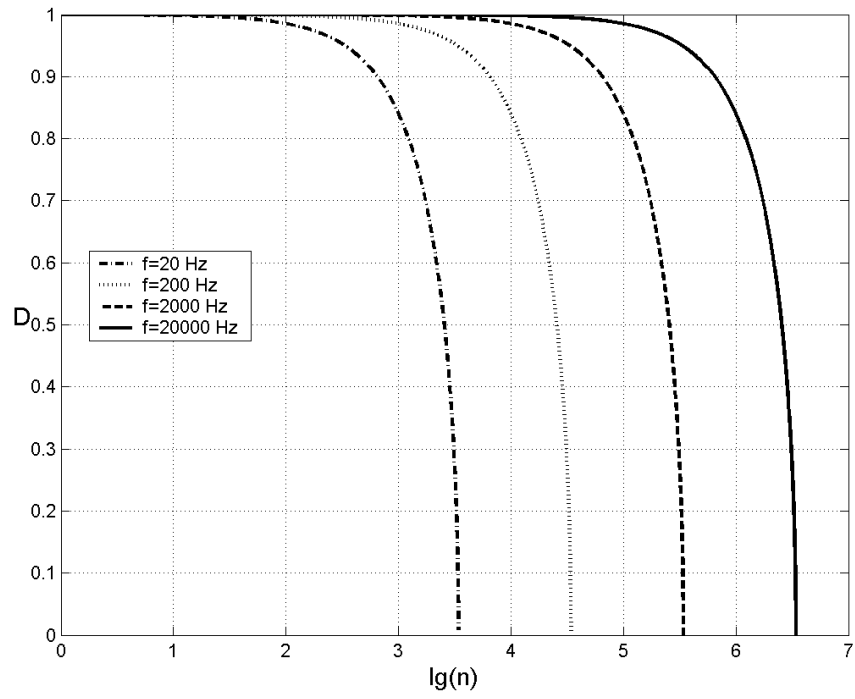

**Figure 2**

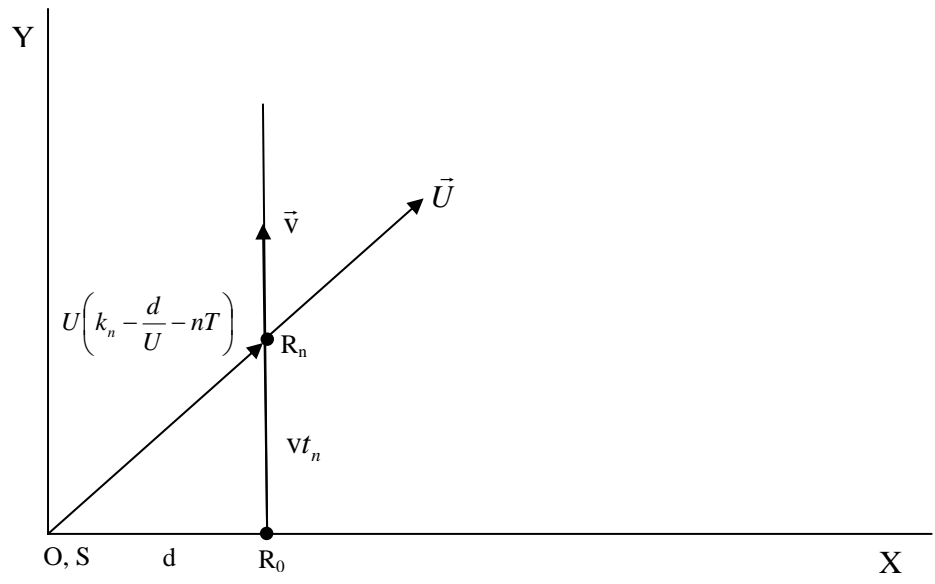

**Figure 3**

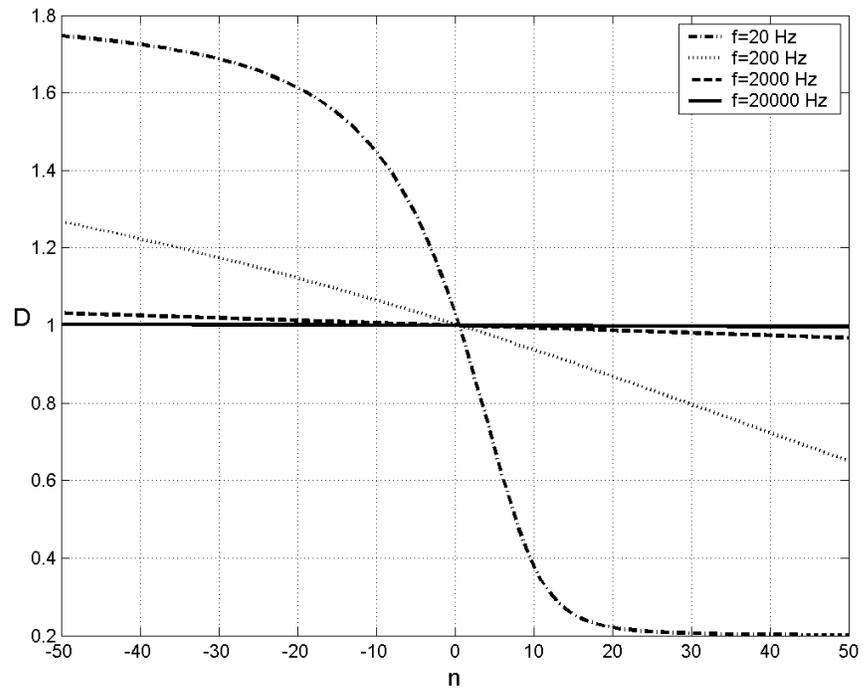

**Figure 4**